\documentstyle[12pt]{article} 
 \def\be{\begin{equation}}
\def\ee{\end{equation}} \def\bea{\begin{eqnarray}}
\def\eea{\end{eqnarray}} 
\begin{document}
\begin{titlepage} \null

\begin{flushright} KOBE-TH-93-08\\ October 1993 \end{flushright}

\vspace{1cm} \begin{center} {\Large\bf Supercurrents on Asymmetric
Orbifolds\par} \vspace{1.5cm} \baselineskip=7mm {\large Yasumasa
IMAMURA\par} \vspace{5mm} {\sl Graduate School of Science and
Technology, Kobe University \\ Rokko-dai, Nada, Kobe 657, Japan \par}
\vspace{1cm}

{\large\bf Abstract} \end{center} \par We study $E_8 \times
E_8$-heterotic string on asymmetric orbifolds associated with
semi-simple simply-laced Lie algebras.  Using the fact that $E_6$-model
allows different twists, we present a new  N=1 space-time supersymmetric
model whose supercurrent appears from twisted sectors but not untwisted
sector.

\end{titlepage}

 \par In searching for ``realistic'' 4-D string theories, there are some
desirable phenomenological requirements. In this paper, we will give
careful consideration to the preservation of N=1 space-time
supersymmetry.  In order to obtain 4-D string theories from 10-D
heterotic string theory\cite{GHMR}, the preservation of N=1 space-time
supersymmetry have played an important role of a guiding principle for
the model selection.  Toroidal orbifold compactification\cite{DHVW} will
be one of the simplest methods to construct the models and has rich
contents.  Orbifolds are defined by dividing tori with discrete
rotations $\theta$. It is convenient to introduce complex coordinates
$Y^i=\frac{1}{\sqrt 2}(X^{2i-1}+iX^{2i})$, ${\bar Y}^i=\frac{1}{\sqrt
2}(X^{2i-1}-iX^{2i})$, (i=1,...,3). In this basis, the rotations are
generated by \be \theta=exp2\pi i[v^1J^{12}+v^2J^{34}+v^3J^{56}], \ee
where $J^{ij}$ are the $SO(6)$ Cartan generators and $v^i$ satisfy
$Nv^i\in {\bf Z}$ and the eigenvalues can be diagonalized of the form
$e^{2\pi i v^i}$.  Then the condition for the preservation of N=1
space-time supersymmetry is given by  \be \pm v^1 \pm v^2 \pm v^3=0,
\label{shift} \ee for some choice of signs\cite{DHVW}.  However, we have
to emphasize that this condition is for symmetric orbifold models but
not for asymmetric ones. We should not confuse the conditions for
symmetric orbifold models with those for asymmetric ones. We will
consider left-right asymmetric orbifold models\cite{Asym} with
left-right different twists $\theta_L$, $\theta_R$.  It seems natural to
study the asymmetric orbifold models because of the asymmetric nature of
the heterotic string theory.  For asymmetric orbifold models, the
condition (\ref{shift}) is not necessarily required for the construction
of N=1 supersymmetric models.  The main purpose of this paper is to show
an example of 4-D N=1 space-time supersymmetric heterotic string
theories without satisfying the condition (\ref{shift}).  \par We start
by describing the ingredients of the heterotic string in the light-cone
gauge. The heterotic string consists of eight left-moving bosonic fields
$X^i_L(\sigma+\tau), (i=1,...,8)$ and extra sixteen fields
$X^I_L(\sigma+\tau), (I=1,...,16)$.  Right-movers include eight bosonic
fields $X^i_R(\sigma-\tau), (i=1,...,8)$ and Neveu-Schwarz-Ramond (NSR)
fermions $\lambda^i(\sigma-\tau)$.  We will introduce bosonized fields
$H^t$ (t=1,...,4) instead of NSR fermions, whose momenta lie on the
weight lattice of SO(8). Then NS sector corresponds to the vector
representation and R sector corresponds to spinor representation of
SO(8) respectively. In order to obtain a 4-D string theory, 22
left-moving and 6 right-moving string coordinates have to be
compactified on tori, orbifolds or some manifolds such as Calabi-Yau
manifolds.  In this paper, we restrict our consideration to orbifold
models, especially on asymmetric orbifold models. \par Orbifolds {\bf O}
are defined as quotient spaces of tori ${\bf T}$ by the action of
discrete rotations $\theta$.  Now we will consider the following models,
\be {\bf O}=\frac{{\bf T}_{L+R}^6}{{\bf Z}_N} \times {\bf T}_L^{E_8
\times E_8}, \ee where ${\bf Z}_N$ is a discrete group and acts as
isometries of the six-dimensional lattice with order $N$; ${\bf Z}_N =
\{ \theta^l, l=0,...,N-1 \}$.  When we divide the left- and right-movers
in the same manner, we will obtain symmetric orbifold
models\cite{model}. If they are not the case, we will obtain asymmetric
orbifold models. In the case of the asymmetric orbifold models, the
twists act on the string coordinates as follows: \be (X_L, X_R)
\rightarrow (\theta_LX_L, \theta_RX_R). \ee In order to construct the
models explicitly, we will restrict the defining tori ${\bf T}^{L+R}_6$
such that they are spanned by root or weight vectors associated with
semi-simple simply-laced Lie algebras G with rank 6.  Let us consider
the momentum lattice $\Gamma_G^{6,6}$. For the sake of modular
invariance of the one-loop partition function, the lattice has to be a
Lorentzian even self-dual lattice\cite{E-N}, \be
\Gamma_G^{6,6}=\{(p_L^i, p_R^i)|\; p_L^i, p_R^i \in \Lambda_W(G) \quad
and \quad p_L^i-p_R^i \in \Lambda_R(G)\}, \ee where $\Lambda_W(G)$
($\Lambda_R(G)$) is the weight (root) lattice of G and the root vectors
are normalized as their squared length is two. Then these choices of the
lattices limit the twists $\theta$ because they have to be automorphisms
of the lattices. G may be given by all possible combinations of the
following groups allowing the multiplicity with total rank 6:  \{SU(2),
SU(3), SU(4), SU(5), SU(6), SU(7), SO(8), SO(10), SO(12), $E_6$ \}. We
select the models such that the automorphisms are inner ones. Moreover,
throughout this paper, we restrict our attention to the ${\bf
Z}_N$-orbifold models whose ${\bf Z}_N$-twists $\theta^l$
($l=1,...,N-1$) leave only the origin fixed.  These allowed inner
automorphisms are classified in ref.\cite{Myhill}. Under these
restrictions, we find that there are four candidates,i.e., \bea
\begin{array}{cccc}  \frac{{\bf T}_{SU(3)^3}}{{\bf Z}_3}&,\quad (v^1,
v^2, v^3) &=& (\frac{1}{3}, \frac{1}{3}, \frac{2}{3}), \\ \frac{{\bf
T}_{E_6}}{{\bf Z}_3}&, \quad (v^1,v^2,v^3) &=& (\frac{1}{3},
\frac{1}{3}, \frac{2}{3}), \\  \frac{{\bf T}_{SU(7)}}{{\bf Z}_7}&, \quad
(v^1, v^2, v^3) &=& (\frac{1}{7}, \frac{2}{7}, \frac{3}{7}), \\
\frac{{\bf T}_{E_6}}{{\bf Z}_9}&, \quad (v^1, v^2, v^3) &=&
(\frac{1}{9},\frac{2}{9}, \frac{5}{9}). \end{array} \label{4-models}
\eea  We attempt to construct 4-D N=1 space-time supersymmetric models
from these models. It is known that for the preservation of 4-D N=1
space-time supersymmetry, $v^i$ must satisfy the condition (\ref{shift})
in the case of symmetric orbifold models.  Therefore, these four models
(\ref{4-models}) will have N=1, 1, 1 and 0 space-time supersymmetry
respectively if they are symmetric orbifold models.  However, for
asymmetric orbifold models, the above condition for the preservation of
4-D N=1 space-time supersymmetry is too restrictive.  In the previous
papers\cite{ISST}, we investigated the supersymmetry of the $E_8 \times
E_8$-heterotic string theories on asymmetric orbifolds of a ``chiral''
type ; ${\bf Z}_N:(X_L, X_R)\rightarrow(X_L, \theta X_R)$ and found that
the four models possess N=2, 4, 4 and 4 space-time supersymmetry
respectively.  In the asymmetric ${\bf Z}_9$-orbifold model, the shift
vector $(v^1, v^2, v^3)=(\frac{1}{9}, \frac{2}{9}, \frac{5}{9})$ does
not satisfy the condition (\ref{shift}) and in fact this shift breaks
the supersymmetry in the untwisted sector.  However, supercurrents
appear from the twisted sectors because the ground state of the
untwisted left-movers has eigenvalue zero with respect to the zero-mode
of the Virasoro operator $L_0$ and (0, 1) conserved currents exist.
These currents play a role of ``twist-untwist intertwining currents''.
The twist-untwist intertwining currents convert untwisted sector to
twisted ones and vice-versa and possess a conformal weight (1, 0) or (0,
1)\cite{Intertwine}. Then the total Hilbert space becomes to possess N=4
supersymmetry.  These four models are interesting in itself, however,
not realistic as they are. We have to construct other class of
asymmetric orbifold models to make the models more realistic with N=1
space-time supersymmetry.  \par We will analyze the $E_8 \times E_8$
heterotic string theory whose extra six-dimensional space is
compactified on an asymmetric orbifold associated with the Lie algebra
$E_6$. We know that this model allows two different twists, ${\bf Z}_3$-
and ${\bf Z}_9$-twists. Let us consider the model whose left-moving
string coordinates are living on a ${\bf Z}_3$-orbifold and the
right-moving ones are on a ${\bf Z}_9$-orbifold. Then the twists act as
follows:  \be (X_L, X_R) \rightarrow (\theta_LX_L, \theta_RX_R), \quad
\theta^3_L={\bf 1}, \quad \theta^9_R={\bf 1}. \ee In the complex basis,
the rotation matrices can always be diagonalized  \bea \theta_L &=&
diag[exp2\pi i(\xi^{(1)}_1, \xi^{(1)}_2, \xi^{(1)}_3)], \\ \theta_R &=&
diag[exp2\pi i(\eta^{(1)}_1, \eta^{(1)}_2, \eta^{(1)}_3)],  \eea and the
left-${\bf Z}_3$ and right-${\bf Z}_9$ model has components
$\xi^{(1)}=(\frac{1}{3}, \frac{1}{3}, \frac{2}{3})$,
$\eta^{(1)}=(\frac{1}{9}, \frac{2}{9}, \frac{5}{9})$.  Because of the
choice of the twists, we can expect that (0, 1) states appear only from
$\theta^3$- and $\theta^6$-twisted sectors and supercurrents appear from
these sectors. Before analyzing these states, we will consider the
one-loop partition function of the $\theta^l$-twisted sector twisted by
$\theta^m$ in order to confirm the consistency of this model. The
partition function is given by  \be Z(\theta^l, \theta^m)=Tr(\theta^m
q^{L_0-D/24}{\bar q}^{{\bar L}_0-D/24}), \quad q=exp(2\pi i \tau), \ee
where the trace is taken over the Hilbert space of the $\theta^l$-sector
and $L_0({\bar L_0})$ is the zero mode of the left-(right-) moving
Virasoro operators. The necessary condition for modular invariance is
then given by invariance under the transformation $\tau \rightarrow
\tau+N/d$, where $d=(l,N)$ is the greatest common divisor of $l$ and
$N$.  In ref.\cite{Vafa}, it is shown that the following condition is
the necessary and sufficient condition for modular invariance and is
called the level-matching condition, \be {N \over
d}(L_0-\bar{L_0})=0\quad mod\; 1. \ee For our purpose, it is sufficient
to check the level-matching for the massless states which concern the
symmetries. The mass formulae for $\theta^l$-twisted sector ($l$=0, for
untwisted sector) are given by \bea \frac{1}{8}(m_L^{(l)})^2 &=& \left\{
\begin{array}{ll} \frac{1}{2}\sum_{i=1}^6(p_L^i)^2
+\frac{1}{2}\sum_{I=1}^{16}(p_L^I)^2+N_L^{(l)} -1, & for\; l=0, 3, 6, \\
\frac{1}{2}\sum_{I=1}^{16}(p_L^I)^2+N_L^{(l)} +E_{L0}^{(l)}-1, & for\; l
\neq 0, 3, 6, \\ \end{array}\right. \\ \frac{1}{8}(m_R^{(l)})^2 &=&
\frac{1}{2}\sum_{i=1}^6(p_R^i)^2 \delta_{l,0}
+\frac{1}{2}\sum_{t=1}^4(p^t+lv^t)^2+N_R^{(l)}
+E_{R0}^{(l)}-\frac{1}{2}, \eea where $N_L^{(l)}$ and $N_R^{(l)}$ are
the number operators. $E_{L0}^{(l)}$ and $E_{R0}^{(l)}$ are
contributions from twisted oscillators to the zero-point energy and are
given by \bea E_{L0}^{(l)}&=&\frac{1}{2}\sum_{i=1}^3
\xi_i^{(l)}(1-\xi_i^{(l)}), \quad \xi_i^{(l)}=l\xi_i^{(1)} \quad mod \;
1, \; 0<\xi_i^{(l)}<1, \\ E_{R0}^{(l)}&=&\frac{1}{2}\sum_{i=1}^6
\eta_i^{(l)}(1-\eta_i^{(l)}), \quad \eta_i^{(l)}=l\eta_i^{(1)} \quad mod
\; 1, \; 0<\eta_i^{(l)}<1. \eea In the untwisted sector, not all
massless states will survive under the generalized GSO projection
introduced for the sake of modular invariance. Since the projection
operator has a form of ${\cal P}=exp 2\pi i (\sum^4_{t=1}{\hat p}^t
v^t)$ with $v^4=0$, we find that the physical massless states are
obtained as follows: \bea |p^t=(0,0,0,\pm 1)>^{unt}_R \otimes \;
\alpha^a_{L-1}|0>_L, \quad a=1, 2, \nonumber \eea where $a$ is a
space-time index in the light-cone gauge. These states correspond to
graviton, antisymmetric background field and a scalar field and \[
|p^t=(0,0,0,\pm 1)>^{unt}_R \otimes  \left\{ \begin{array}{rl}
&\alpha^{I}_{L-1}|0>_L \\ &|(p_L^I)^2=2>_L \\ &\alpha^{i}_{L-1}|0>_L \\
&|(p_L^i)^2=2>_L \end{array}\right. \]  correspond to Yang-Mills fields.
It should be emphasized that there is no superpartner in this sector.
Let us consider the twisted sectors. Because of our choice of the model,
the contributions from the twisted oscillators to the zero-point energy
are given by $(E_{L0}^{(l)}, E_{R0}^{(l)}) = (\frac{1}{3},
\frac{7}{27}), (l \neq 3, 6)$ and find that there is no massless
superpartner in these sectors since massless states cannot have
space-time indices. However, for the $\theta^3$- and $\theta^6$-sectors,
the orbifold becomes chiral type with $(E_{L0}^{(l)}, E_{R0}^{(l)}) =
(0, \frac{1}{3}), (l=3, 6)$ and the massless states are given as
follows:

 $\theta^3$-sector: \bea |p^t=(-\frac{1}{2}, -\frac{1}{2}, -\frac{3}{2},
\frac{1}{2})>^{\theta^3}_R &\otimes& \alpha^a_{L-1}|0>_L, \quad a=1, 2
\nonumber \\ |p^t=(-\frac{1}{2}, -\frac{1}{2}, -\frac{3}{2},
\frac{1}{2})>^{\theta^3}_R &\otimes&  \left\{ \begin{array}{rl}
&\alpha^{I}_{L-1}|0>_L \\ &|(p_L^I)^2=2>_L \\ &\alpha^{i}_{L-1}|0>_L \\
&|(p_L^i)^2=2>_L \end{array}\right.  \nonumber \eea

 $\theta^6$-sector:  \bea |p^t=(-\frac{1}{2}, -\frac{3}{2},
-\frac{7}{2}, -\frac{1}{2})>^{\theta^6}_R &\otimes& \alpha^a_{L-1}|0>_L,
\quad a=1,2 \nonumber \\ |p^t=(-\frac{1}{2}, -\frac{3}{2}, -\frac{7}{2},
-\frac{1}{2})>^{\theta^6}_R &\otimes&  \left\{ \begin{array}{rl}
&\alpha^{I}_{L-1}|0>_L \\ &|(p_L^I)^2=2>_L \\ &\alpha^{i}_{L-1}|0>_L \\
&|(p_L^i)^2=2>_L \end{array}\right.  \nonumber \eea These are
superpartners of gravity and Yang-Mills fields. Together with these
states, we can conclude that this model has N=1 supergravity and super
Yang-Mills multiplet and realizes the N=1 space-time supersymmetry.
Here, we have to mention that the degeneracy of the ground states in the
$\theta^3$- and $\theta^6$-twisted sectors is given by \be
n=\frac{\sqrt{det(1-\theta^l_R)}}{V_{\Lambda_R (E_6)}}=3, \hspace{1cm}
l=3,6, \ee where $V_{\Lambda_R (E_6)}$ is a volume of the unit cell of
$\Lambda_R (E_6)$. In order to evaluate the physical states, a detailed
analysis of the one-loop partition function is required.  For ${\bf
Z}_N$-orbifolds with non-prime N, a further non-trivial projection
appears.  \bea Z(\tau)_{\theta^3-sector} &=& \frac{1}{9}\sum_{m=0}^8
Z(\theta^3, \theta^m; \tau), \\ Z(\tau)_{\theta^6-sector} &=&
\frac{1}{9}\sum_{m=0}^8 Z(\theta^6, \theta^m; \tau). \eea Expanding
these partition functions in powers of $q=e^{2\pi i \tau}$ and ${\bar
q}=e^{2\pi i {\bar \tau}}$, we can show that one of the three degeneracy
states survives.

The supercurrents are constructed from fermion vertex
operators\cite{FVO}. In the covariant formalism, they are constructed by
introducing a spin field $e^{-\phi/2}$ and five free scalar fields
$H^\nu$ ($\nu$=1,...,5) representing the NSR fermions through
bosonization. The fermion vertex operators for the untwisted sector are
given by  \be V_{-\frac{1}{2}}=e^{-\phi/2}e^{i\alpha_s H}e^{ikX},  \ee
and for the $\theta^l$-twisted sector \be
V_{-\frac{1}{2}}=e^{-\phi/2}e^{i(\alpha_s+v^{(l)}) H}e^{ikX}\Lambda, \ee
where $e^{-\phi/2}$ is a spin field with conformal dimension
$\frac{3}{8}$ for the $(\beta, \gamma)$ superconformal ghost system.
$\alpha_s$ are spinorial weights of SO(10), $\alpha_s=(\pm
\frac{1}{2},..., \pm \frac{1}{2})$ with even number of $+$ signs and
$v^{(l)}=(0, 0, v_1^{(l)}, v_2^{(l)}, v_3^{(l)}), v_i^{(l)}=lv_i \; mod
\; 1, \; 0<v_i^{(l)}<1$. $k^{\mu}$ is a 4-dimensional momentum.  Note
that we replace SO(8) by SO(10) in order to describe in the covariant
formalism. $\Lambda$ is a twisted field which creates twisted vacuum out
of the untwisted one and is defined by \be \Lambda(0)|0> = |\Lambda>.
\ee The conformal dimension of the twist field $\Lambda$ is easily
calculated by \be \frac{1}{2}\sum_{i=1}^{3}\eta^{(l)}_i(1-\eta^{(l)}_i).
\ee If the conformal weight of the supercurrents is (1, 0) and these
currents will play a role of twist-untwist intertwining currents. It is
convenient to rewrite the fermion vertex operators into the form that
4-D space-time supersymmetry is manifest:  \bea
V_{-\frac{1}{2}}^{\alpha}(\bar z)&=&e^{-\phi/2}S^{\alpha}\Sigma(\bar z),
\label{TSC1}\\ V_{-\frac{1}{2}}^{\dot \alpha}(\bar
z)&=&e^{-\phi/2}S^{\dot \alpha}\Sigma^{\dag}(\bar z), \label{TSC2}\eea
where $S^\alpha$ are the spin fields given as exponentials of two free
bosons $H^{1, 2}$ with conformal dimension $\frac{1}{4}$ and
$\Sigma$($\Sigma^{\dag}$) are dimension $\frac{3}{8}$ fields constructed
from exponentials of three free bosons $H^{3,4,5}$ and $\Lambda$.  Then
the 4-D supersymmetry charges are given by $Q_{\alpha}=\oint d{\bar z}
V^{\alpha}_{-1/2}(\bar z), Q_{\dot \alpha}=\oint d{\bar z} V^{\dot
\alpha}_{-1/2}(\bar z) ,\; \alpha=(\pm \frac{1}{2}, \pm \frac{1}{2}),
{\dot \alpha}=(\pm \frac{1}{2}, \mp \frac{1}{2})$.  \par We gave a new
4-D N=1 space-time supersymmetric model compactified on an asymmetric
orbifold associated with the Lie algebra $E_6$. Here, we restricted the
models whose ${\bf Z}_N$-twists $\theta^l$ ($l=1,...,N-1$) did not have
fixed direction.  Under this condition, ${\bf Z}_3$- and ${\bf
Z}_9$-twists are allowed for the $E_6$-model. We divided the bosonic
left-moving string coordinates with the ${\bf Z}_3$-twist and the
right-moving superstring coordinates with the ${\bf Z}_9$-twist.   The
${\bf Z}_9$-twist does not satisfy the condition for the preservation of
N=1 space-time supersymmetry $\pm v^1 \pm v^2\pm v^3=0$ for any choice
of signs.   We found that this model has no fermionic massless state in
the untwisted sector.  However, supercurrents appear from $\theta^3$-
and $\theta^6$-twisted sectors and N=1 space-time supersymmetry is
realized in the total Hilbert space.  Then the supercurrent is given by
the form of eqs.(\ref{TSC1}) (\ref{TSC2}) and plays a role of
intertwiner between untwisted and twisted sectors.  It is remarkable
that all matter fields appear from the twisted sectors. \par There may
be some other examples which possess 4-D N=1 supersymmetry in the same
way. In this paper we did not treat the gauge sector and we thought
non-embedding models.  The standard embedding, non-standard embedding
and Wilson-line mechanism will be able to consider so as to satisfy
modular invariance for these models.  \par The author is grateful to Dr.
M. Sakamoto for suggestive discussions and for careful reading of the
manuscript.

 \end{document}